\documentclass[showpacs,prl,aps,superscriptaddress,twocolumn]{revtex4-1}

\pdfoutput=1

\usepackage{amsmath,amsfonts,amssymb,bm,graphicx,hyperref,color}
\newcommand{\C}{\mathcal{C}}
\newcommand{\W}{\mathbb{W}}
\newcommand{\llangle}{\langle\!\langle}
\newcommand{\rrangle}{\rangle\!\rangle}

\begin{document}
\title{Trajectory phase transitions, Lee--Yang zeros, and high-order cumulants \\in full counting statistics}

\author{Christian Flindt}
\affiliation{D\'epartement de Physique Th\'eorique, Universit\'e de Gen\`eve, 1211 Gen\`eve, Switzerland}
\author{Juan P.\ Garrahan}
\affiliation{School of Physics and Astronomy, University of Nottingham, Nottingham, NG7 2RD, United Kingdom}
\date{\today}

\begin{abstract}
We investigate Lee--Yang zeros of generating functions of dynamical observables and establish a general relation between phase transitions in ensembles of trajectories of stochastic many-body systems and the time evolution of high-order cumulants of such observables.  This connects dynamical free-energies for full counting statistics in the long-time limit, which can be obtained via large-deviation methods and whose singularities indicate dynamical phase transitions, to observables that are directly accessible in simulation and experiment.  As an illustration we consider facilitated spin models of glasses and show that from the short-time behavior of high-order cumulants it is possible to infer the existence and location of dynamical or ``space-time'' transitions in these systems.
\end{abstract}



\maketitle

\emph{Introduction}.--- Phase transitions are a central topic in the statistical mechanics of equilibrium and non-equilibrium systems. In problems with physically meaningful interactions, phase transitions occur in the limit of large system size.  For dynamical phase transitions, this also implies the limit of long times.  In experiment or simulation of systems with complex dynamics, however, often only the short-time dynamics can be probed, making it difficult to investigate dynamical transitions.  Furthermore, such non-equilibrium transitions may be driven by ``counting'' fields \cite{Lecomte2007,Garrahan2007,*Garrahan2009,Hedges2009,*Pitard2011,Giardina2011,Gorissen2009,*Jack2010,*Elmatad2010,Chernyak2010,*Hurtado2011,*Dickson2011,*Monthus2011,Garrahan2010,*Budini2011,*Ates2012,levkivskyi2009,*karzig2010,*ivanov2010,*Alvarez2010,*Li2011} which can be hard to relate to physically accessible parameters.  In this Letter we provide a potential resolution to these problems by establishing a connection between phase transitions in ensembles of long-time dynamical trajectories of classical stochastic many-body systems \cite{Lecomte2007,Garrahan2007,*Garrahan2009,Hedges2009,Giardina2011,Gorissen2009} and the dynamics of physical observables at short times \cite{Flindt2009,Flindt2010,Kambly2011}.

Figure 1 illustrates our approach and results.  Panel~(a) shows a dynamical trajectory of a simple lattice system which displays complex dynamics, in this example the one-dimensional East model of a glass former \cite{[{For a review see }]Ritort2003}. Facilitated models such as the East model show pronounced dynamical spatial fluctuations \cite{Garrahan2002} (a phenomenon characteristic of glasses known as dynamical heterogeneity; for reviews see Refs.\ \cite{Ediger2000,Berthier2011,Chandler2010}).  These large spatio-temporal fluctuations give rise to fat tails \cite{Merolle2005} in the full counting statistics (FCS) \cite{Levitov1993,*Bagrets2003,*Pilgram2003,*Flindt2008,*Esposito2009} of time-extensive dynamical observables.  This is shown in Fig.~1(b) for the dynamical {\em activity} $k\equiv K/t$ per unit time of the East model. The dynamical activity $K$ is the number of configuration changes in a trajectory \cite{Lecomte2007,Garrahan2007,Baiesi2009,Giardina2011}.  Associated with the distribution $P(K,t)$ is the moment generating function (MGF) $Z(s,t) \equiv \sum_K e^{-sK} P(K,t)$, which at long times $t \to \infty$ has a large-deviation (LD) form, $Z(s,t) \propto \exp{\{t \theta(s)\}}$~ \cite{Lecomte2007,Garrahan2007,*Garrahan2009,Hedges2009,Giardina2011,Gorissen2009}.  The LD function $-\theta(s)$ is a dynamical free-energy for the counting process.  Its analytic properties carry information about the phase behavior of ensembles of trajectories.

\begin{figure}[t!]
\includegraphics[width=0.9\columnwidth]{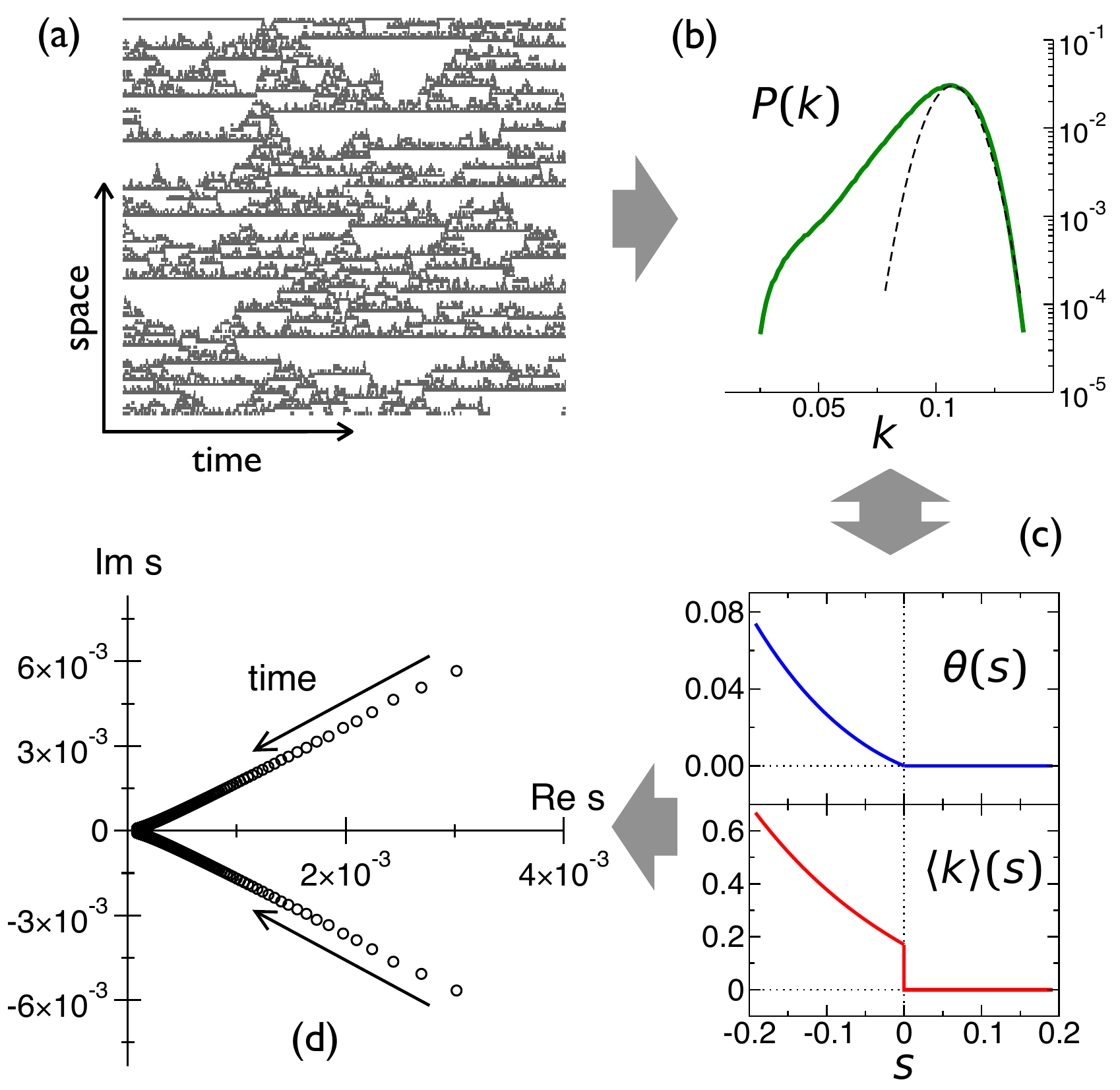}
\caption{(color online). Dynamical phase transitions and Lee--Yang zeros. (a) Trajectory of the one-dimensional East model, showing the state of up/down (black/white) spins on the lattice in time (with temperature $T=0.8$, $N=150$ lattices sites, and $t_{\rm max}=1000$ time steps); dynamic heterogeneity is evident in the ``space-time bubbles'' of the trajectory \cite{Garrahan2002}. (b) Probability $P(k,t)$ of the activity $k\equiv K/t$ per unit time (green/full curve); dashed line is a Gaussian distribution with same mean and variance. (c) The LD function $\theta(s)$ is singular at $s_c=0$ (blue/top) where the average activity is discontinuous (red/bottom), indicative of a first-order dynamical transition \cite{Garrahan2007}.  (d) Lee--Yang zeros of the MGF in the complex-$s$ plane, extracted from the cumulants of $K$, allow us to extrapolate $s_c$ from short-time observables.
\label{Fig1}}
\end{figure}

In the East model example $\theta(s)$ has a first-order singularity at $s_c=0$, Fig.~1(c), which indicates that dynamics takes place at the coexistence of two dynamical or ``space-time'' phases, an active phase with $\langle k \rangle(s) \equiv -\partial_s \theta(s)> 0$ for $t\rightarrow\infty$ (the equilibrium phase where relaxation is possible) and an inactive phase with $\langle k \rangle(s) = 0$ (the dynamical ``glass'' phase) \cite{Garrahan2007}.
The variable $s$ driving the transition is a ``counting'' field which biases the trajectory ensemble from the actual dynamical one at $s=0$, but whose connection to physically controllable parameters can be hard to establish.
Similar trajectory phase transitions are observed in other classical and quantum systems with complex dynamics \cite{Giardina2011,Hedges2009,*Pitard2011,Gorissen2009,*Jack2010,*Elmatad2010,
Chernyak2010,*Hurtado2011,*Dickson2011,*Monthus2011,Garrahan2010,*Budini2011,*Ates2012,levkivskyi2009,*karzig2010,*ivanov2010,*Alvarez2010,*Li2011}.

Here we demonstrate that it is possible to infer the existence and location of singularities of $\theta(s)$, indicative of phase transitions in the space of long-time trajectories, from short-time observables at $s=0$.  Specifically, we show that: (i) from a dynamical version of the Lee--Yang theorem \cite{Lee1952,*Yang1952}, zeros of the MGF in the complex-$s$ plane at finite $t$ will move to the real-$s$ line in the limit of $t \to \infty$ if there are any singularities in $\theta(s)$; and (ii) these zeros can be obtained from the short-time and finite-size behavior of cumulants \cite{Flindt2009,Flindt2010,Kambly2011} of dynamic observables such as the activity.  Figure 1(d) illustrates this result for the East model: the $s_c = 0$ singularity of the thermodynamic and long-time limit can be extrapolated from the leading Lee--Yang zeros extracted from short-time cumulant dynamics.  This offers the possibility of studying trajectory phase transitions in FCS via observables that are directly accessible in simulation and experiment.

\emph{Formalism.}---
For concreteness we consider stochastic processes described by the Master equation \cite{Gardiner1986}
\begin{equation}
\partial_t P(\mathcal{\C},t)=-r(\C)P(\C,t)+\sum_{\C'}W(\C'\rightarrow\C)P(\C',t).
\label{eq:masterequation}
\end{equation}
Here, $P(\mathcal{\C},t)$ is the probability that the system is in the configuration $\mathcal{\C}$ at time $t$. The transition rate from configuration $\mathcal{\C}'$ to $\mathcal{\C}$ is denoted as $W(\C'\rightarrow\C)$ and $r(\C)=\sum_{\C'}W(\C\rightarrow\C')$ is the total escape rate from $\C$. By definition $W(\C\rightarrow\C)=0$. Equation \eqref{eq:masterequation} can be written in the convenient matrix notation $\partial_t |P(t)\rangle=\W |P(t)\rangle$, where the matrix $\W$ is defined as
\begin{equation}
\W(\C,\C') \equiv W(\C'\rightarrow\C)-r(\C)\delta_{\C,\C'} ,
\label{W}
\end{equation}
and the vector $|P(t)\rangle$ contains the probabilities $P(\mathcal{\C},t)$'s.

We classify trajectories according to their dynamical activity $K$---the total number of spin-flips in the case of spin models considered here \cite{Lecomte2007,Garrahan2007}. (Similar arguments can be applied to analyze ensembles of trajectories classified by other time-extensive dynamic observables, see e.~g.~Refs.\ \cite{Lecomte2007,Garrahan2007,Hedges2009}). The probability that the system is in configuration $\C$ at time $t$, having changed configuration $K$ times, is denoted as $P(\C | K,t)$.  Then $P(K,t)=\sum_\C P(\C | K,t)$ and $Z(s,t)=\sum_\C P(\C,s,t)$, where $P(\C,s,t)=\sum_K P(\C | K,t) e^{-sK}$ \cite{Lecomte2007,Garrahan2009}. The corresponding vector $|P(s,t)\rangle$ obeys $\partial_t |P(s,t)\rangle=\W_s |P(s,t)\rangle$, where the generalized Master operator is \cite{Lecomte2007,Garrahan2009}
\begin{equation}
\W_s(\C,\C') \equiv e^{-s} W(\C'\rightarrow\C)-r(\C)\delta_{\C,\C'} .
\label{Ws}
\end{equation}
Formally, the solution to Eq.\ (\ref{Ws}) is $ |P(s,t)\rangle=e^{\W_s t} |P(0)\rangle$, assuming for instance that the initial state $|P(0)\rangle$ is the equilibrium distribution defined by $\W_{s=0} |P(0)\rangle=0$. By using the ``flat'' state, $\langle -| \equiv (1,\ldots,1)$, we can express the MGF as $Z(s,t)=\langle - |P(s,t)\rangle=\langle -| e^{\W_s t} |P(0)\rangle=\sum_j c_j(s)e^{\lambda_j(s)t}$ in terms of the eigenvalues $\lambda_j(s)$ of $\W_s$ and corresponding expansion coefficients $c_j(s)$. The cumulant generating function (CGF) is defined in terms of the MGF as $\Theta(s,t) \equiv \log Z(s,t)$, which delivers the cumulants of $K$ by differentiation with respect to the counting variable $s$ at $s=0$,
\begin{equation}
\llangle K^n\rrangle (t) = (-1)^n \partial_s^n \Theta(s,t)|_{s\rightarrow0}.
\label{eq:cumulantsdef}
\end{equation}
At long times the MGF function becomes exponential in time \cite{Lecomte2007}; its rate of change is determined by the eigenvalue with the largest real-part, such that $\Theta(s,t)\rightarrow t \theta(s)$, where $\theta(s) \equiv \max[\lambda_j(s)]$ is the LD function.

\emph{Singularities and dynamical transitions.}--- Fluctuations in the dynamical system can be understood from the analytic properties of $\theta(s)$. For example, a first-order dynamical phase transition corresponds to singularities in $\theta(s)$ so that its first derivative is discontinuous \cite{Garrahan2007}, see Fig.~1(c).  This occurs at a real $s=s_c$ where the two largest eigenvalues of $\W_s$ become degenerate, $\lambda_0(s_c)= \lambda_1(s_c)$.
As a central result of this work, we show below how such dynamical phase transitions, occurring in the long-time limit, can be inferred from the high-order cumulants of $K$ at \emph{finite} times and at $s=0$, i.~e.\ evolving under the unbiased dynamics.

To this end we consider the zeros of the MGF in the vicinity of the transition value, $s\simeq s_c$,  where the two largest eigenvalues are nearly degenerate $\lambda_0(s)\simeq \lambda_1(s)$ and we may write $Z(s,t)\simeq c_0(s)e^{\lambda_0(s)t}+c_1(s)e^{\lambda_1(s)t}$. The zeros of the MCF are determined by the equations $\lambda_0(s)=\lambda_1(s)+[\log{c_1(s)/c_0(s)}+i\pi (2m+1)]/t$ for integer $m$. In the long-time limit, these equations all reduce to $\lambda_0(s)=\lambda_1(s)$, and thus with increasing time all zeros $s_j(t)$ move towards the transition value $s_c$ on the real-axis. (At finite times, the zeros must be complex, since $Z(s,t)>0$ for real $s$.) This is in essence the theory of phase transitions of Lee and Yang \cite{Yang1952}, here applied to dynamical systems \cite{[{For other dynamical applications of Lee--Yang ideas see e.~g.\ }]Blythe2002,*Bena2005}. Accordingly, we refer to the (time-dependent) zeros $s_j(t)$ of the MGF as Lee--Yang zeros.

\emph{High-order cumulants and Lee--Yang zeros.}--- The motion of the Lee--Yang zeros in the complex plane can be inferred from the high-order cumulants of $K$. Importantly, the zeros of the MGF correspond to logarithmic singularities of the CGF which determine the high-order derivatives of the CGF (the cumulants) according to Darboux's theorem \cite{Dingle1973,*Berry2005}. Writing the MGF in terms of the Lee--Yang zeros as $Z(s,t)=\prod_j[s_j(t)-s]/s_j(t)$, where $Z(0,t)=1$ reflects the normalization $\sum_KP(K,t)=1$ at all times, the CGF becomes $\Theta(s,t)=\sum_j (\log[s_j(t)-s]-\log[s_j(t)])$. The Lee--Yang zeros come in complex-conjugate pairs, since the MGF is real for real $s$. Combined with Eq.\ \eqref{eq:cumulantsdef} we readily find
\cite{Flindt2009,Flindt2010,Kambly2011}
\begin{equation}
\llangle K^n\rrangle(t) = (-1)^{(n-1)}(n-1)!\sum_j\frac{e^{-i n\arg[s_j(t)]}}{|s_j(t)|^n}.
\label{eq:darboux}
\end{equation}
This result shows that higher-order cumulants generically grow as the factorial of the cumulant order $n$, and oscillate as a function of any parameter that changes the complex argument $\arg\{s_j(t)\}$ \cite{Flindt2009}. This behavior has been observed experimentally \cite{Flindt2009,Fricke2010}.  For large $n$, the sum is dominated by the pair $s_0(t)$ and $s_0^*(t)$ of zeros closest to $s=0$, and the expression further simplifies to \cite{Flindt2009,Flindt2010,Kambly2011,Bhalerao2003}
\begin{equation}
\llangle K^n\rrangle(t) \simeq (-1)^{(n-1)}(n-1)!\,\frac{2\cos[n\arg{s_0(t)}]}{|s_0(t)|^n}.
\label{eq:onepoleapprox}
\end{equation}
We can solve this simple relation for $s_0$, given the ratios of cumulants $\kappa_n^{(\pm)}(t)\equiv \llangle K^{n\pm1}\rrangle(t)/\llangle K^n\rrangle(t)$. We then obtain the matrix equation
\begin{equation}
\left[
\begin{array}{cc}
1 & -\frac{\kappa_{n}^{(+)}}{n}  \vspace{.05 cm} \\
1 & -\frac{\kappa_{n+1}^{(+)}}{n+1}\\
\end{array}
\right]
\cdot
\left[
\begin{array}{c}
-(s_0+s^*_0) \\
|s_0|^2\\
\end{array}
\right]
=
\left[
\begin{array}{c}
(n-1)\kappa_{n}^{(-)}   \vspace{.25 cm}\\
n\kappa_{n+1}^{(-)}\\
\end{array}
\right]
\label{eq:cumuextract}
\end{equation}
which directly yields $s_0(t)$ from four consecutive cumulants \cite{Zamastil2005,Flindt2010,Kambly2011}. We now employ this method to investigate dynamical phase transitions in kinetically constrained models of glass formers.

\emph{Dynamical transitions in facilitated glass models.}---
As an example of how the ideas above can be applied, we study trajectory transitions \cite{Garrahan2007} in facilitated spin models of glasses  \cite{Ritort2003}.  For simplicity we consider one-dimensional models, defined in terms of binary variables $n_i=0,1$, where $i=1,\ldots,N$ denote sites on a chain.  The energy function is $E= J \sum_i n_i$, and all interactions emerge via kinetic constrains, which stipulate that a site $i$ changes with a rate that is determined by the state of its nearest neighbors $i\pm1$ \cite{Ritort2003}. Concretely, we focus on the Fredrickson--Andersen (FA) model \cite{Fredrickson1984} and on the East model \cite{Jackle1991}.  In the FA model a site can only change if either of its nearest neighbors is in the up state, i.~e.\ the transitions $11 \rightarrow 10$ and $11 \rightarrow 01$ occur with rate $1$, $11 \leftarrow 10$ and $11 \leftarrow 01$ with rate $e^{-J/T}$, but $010 \leftrightharpoons 000$ are not allowed. In the East model facilitation is via the left neighbor only, so that $11 \rightarrow 10$ and $11 \leftarrow 10$ occur with rates $1$ and $e^{-J/T}$, respectively, but $010 \leftrightharpoons 000$ and $011 \leftrightharpoons 001$ are not allowed. At low $T$, there is a conflict between lowering the energy and having enough excited spins to evolve dynamically, which gives rise to glassy slow-down and dynamical heterogeneity \cite{Garrahan2002,Ritort2003} in these systems; the East model in particular seems to capture the basic physics of glassy dynamical arrest \cite{Chandler2010}.

\begin{figure}
\includegraphics[width=0.98\columnwidth]{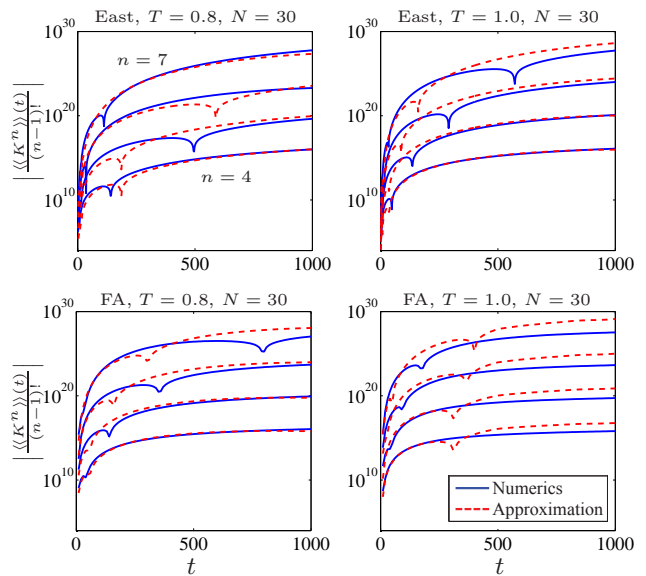}
\caption{(color online). High-order cumulants of the activity in facilitated models of glass formers. The upper (lower) panels show the time-evolution of the cumulants of order $n=4-7$ for the East (FA) model with $N=30$ sites at $T=0.8$ (left) and $T=1.0$ (right). Simulations are shown with full lines, while dashed lines indicate the approximation Eq.\ (\ref{eq:onepoleapprox}) based on the closest pair of Lee--Yang zeros which are extracted from the numerical data using Eq.\ (\ref{eq:cumuextract}). The motion of the closest pair of Lee--Yang zeros corresponding to the East model at $T=0.8$ is shown in Fig.~\ref{Fig1}(d).
\label{Fig2}}
\end{figure}

\emph{Results.}--- Figure \ref{Fig2} shows our numerical simulations for the high-order cumulants of the activity $K$ as functions of time for the East and FA models (full lines). The cumulants grow dramatically with the cumulant order and oscillate as functions of time (the absolute value is shown on a logarithmic scale, such that downwards-pointing spikes on the curves correspond to the cumulants crossing zero). This is due to the Lee--Yang zeros $s_j(t)$ approaching the transition value at $s_c=0$ according to Eq.~(\ref{eq:darboux}), causing the large growth of the cumulants. Initially, $P(K,t=0)=\delta_{K,0}$ and all cumulants of the activity are zero, implying that the Lee--Yang zeros are infinitely far from $s_c=0$ and $1/|s_j(t=0)|=0$. At very short time, where $P(K=0,t)\approx 1>P(1,t)\gg P(2,t)\gg \ldots$, the leading pair of Lee--Yang zeros are determined by the equation $Z(s,t)\simeq P(0,t)+P(1,t)e^{-s}=0$ with solutions $s_0(t),\,s_0^*(t)=-\log\{P(0,t)/P(1,t)\}\pm i\pi$. Thus, to begin with the Lee-Yang zeros move along the lines $\pm i\pi$ from $-\infty\pm i\pi$, before approaching $s_c=0$. We now use Eq.~(\ref{eq:cumuextract}) to deduce the motion of the leading Lee--Yang zeros from the numerical data.

Figure \ref{Fig1}(d) shows the leading pair of Lee--Yang zeros, $s_0(t)$ and $s^*_0(t)$, for the East model with $N=30$ sites at temperature $T=0.8$ as they move towards the first-order transition point at $s_c=0$. To validate the extraction of the leading Lee--Yang zeros from the cumulants of the activity using Eq.\ (\ref{eq:cumuextract}), we plug the solution $s_0(t)$ back into Eq.\ (\ref{eq:onepoleapprox}) and compare the result with the numerical data. In Fig.~\ref{Fig2} we show the numerical results (full lines) together with the approximation in Eq.\ (\ref{eq:onepoleapprox}) based on the extracted pair of Lee--Yang zeros (dashed line). The figure corroborates that we indeed are extracting the leading pair of Lee--Yang zeros. Some deviations, in particular at long times, are observed as the second pair of Lee--Yang zeros also come close to $s=0$ and start contributing significantly to the sum in Eq.\ (\ref{eq:darboux}). Since the second pair of Lee-Yang zeros is not included in Eq.\ (\ref{eq:onepoleapprox}), a shift in the frequency of the oscillations as a function of time is also observed. If needed, the accuracy of the method can be improved by using higher cumulants \cite{Zamastil2005,Flindt2010,Kambly2011}.

\begin{figure}
\includegraphics[width=0.98\columnwidth]{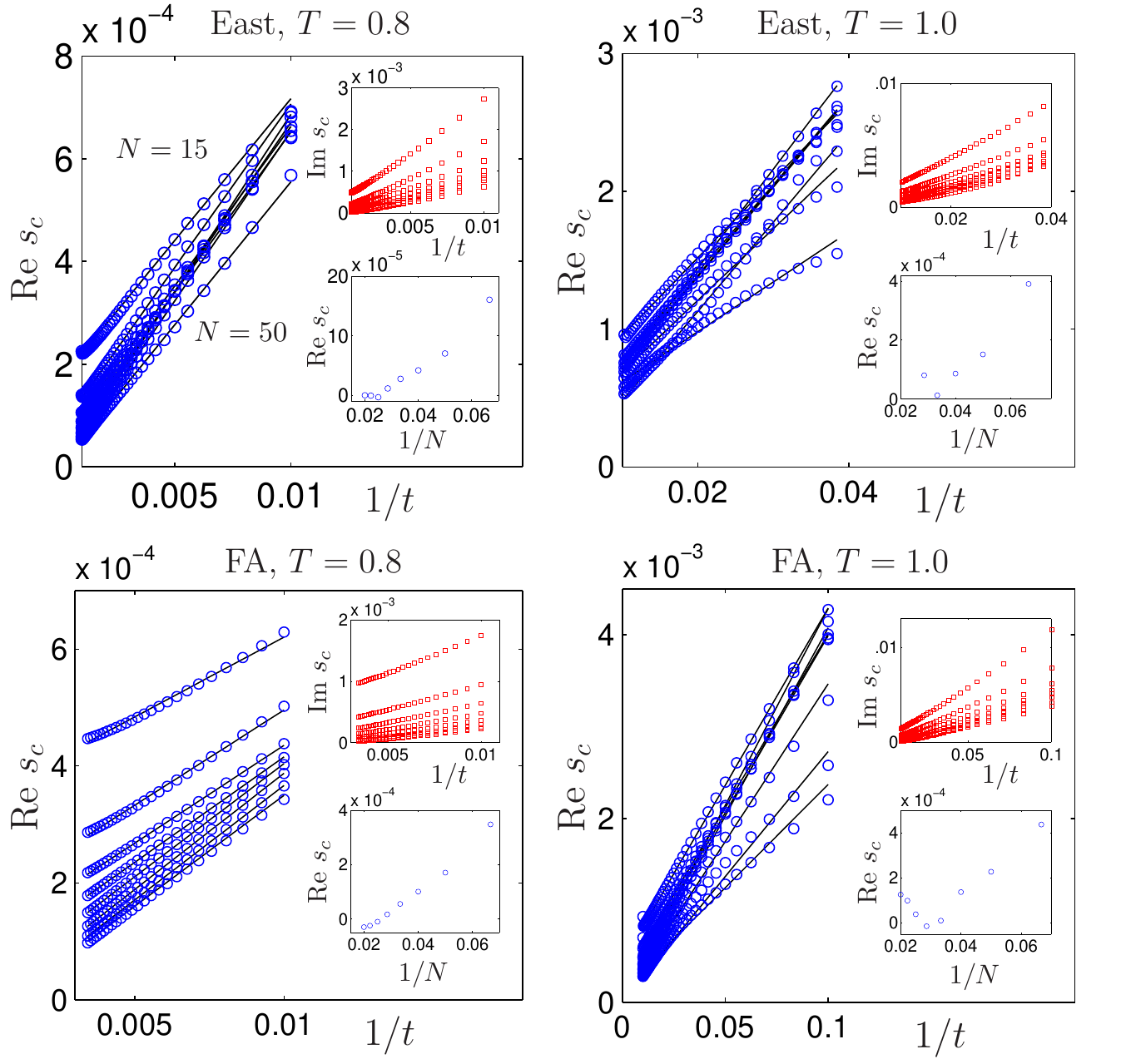}
\caption{(color online). Finite-size scaling of the first-order transition point.  Upper (lower) panels show the real-part of $s_c(t)$, extracted from the cumulants for the East (FA) model with $N=15-50$ sites at $T=0.8$ (left) and $T=1.0$ (right). Numerical values are shown with circles, while linear fits based on Eq.\ (\ref{eq:scaling}) are shown with full lines. Upper insets show the imaginary part of $s_c(t)$ as a function of the inverse time $1/t$. Lower insets show the extrapolated long-time limit $s_c^{(\infty)}(N)$ as a function of the inverse system size $1/N$.
\label{Fig3}}
\end{figure}

In Fig.~\ref{Fig3} we analyze the finite-size scaling of the transition point $s_c$. The real-part of the transition point is predicted to scale as \cite{Bodineau2012a,*Bodineau2012b}
\begin{equation}
\mathrm{Re}[s_c(t)]\simeq\alpha/t+s_c^{(\infty)}(N),
\label{eq:scaling}
\end{equation}
where the coefficient $\alpha$ depends on the temperature $T$, and $s_c^{(\infty)}(N)\propto 1/N$ is the long-time value, which for the East and FA models should approach $s_c=0$ in the limit $N\rightarrow\infty$. Our numerical results for the East and FA models confirm the predicted scaling behavior. For each system size in the range $N=15$ to $50$, we find an approximately linear dependence on the inverse time $1/t$, allowing us to extrapolate the values of $s_c^{(\infty)}(N)$ in the $t\rightarrow\infty$ limit. We also verify that the imaginary part of $s_c$ approaches zero in the long-time/large-system limit, see upper insets. In the lower insets, we show the extrapolated values of $s_c^{(\infty)}(N)$ as a function of the inverse system size $1/N$. These results show that the value $s_c=0$ is approached in the large-system-size limit. Some deviations are seen for the larger systems as we reach the limits of the numerical accuracy of our method. Our results show that it is possible to infer the existence and location of dynamical singular points, which are indicative of phase transitions in the space of long-time trajectories, from high-order short-time cumulants at $s=0$. Our method can also be extended to systems where the transition point on the real-$s$ line is at $s_c \neq 0$ \cite{Gorissen2009}.

\emph{Conclusions}.--- We have investigated the Lee-Yang zeros of generating functions of dynamical observables and demonstrated how singularities in the long-time limit, indicative of dynamical phase transitions, can be inferred from the short-time dynamics of high-order cumulants in finite-size systems. We hope that our approach may facilitate theoretical and experimental studies of trajectory phase transitions in stochastic many-body systems. An important task to address in future work is to apply similar ideas to dynamical phase transitions in quantum many-body systems \cite{Garrahan2010,levkivskyi2009}.

\emph{Acknowledgments}.--- The work was supported by Swiss NSF, by EPSRC Grant no.\ EP/I017828/1 and Leverhulme Trust grant no.\ F/00114/BG.

\end{document}